\documentclass{iopart}%

\def\binom#1#2{{#1 \choose #2}}

\begin{document}

\title[Geometry of the standard model]{A geometric basis for the
standard-model\\gauge group}
\author{Greg Trayling and W E Baylis}
\address{Department of Physics, University of Windsor, Windsor, Ont., Canada
N9B 3P4\\\medskip E-mail: traylin@uwindsor.ca,
baylis@uwindsor.ca\\\medskip Received 5 October 2000, in final
form 15 January 2001.\\Scheduled for publication in \emph{J Phys
A: Math Gen} April 2001}

\begin{abstract}
A geometric approach to the standard model in terms of the Clifford algebra
$C\!\ell_{7}$ is advanced. A key feature of the model is its use of an
algebraic spinor for one generation of leptons and quarks. Spinor
transformations separate into left-sided (``exterior'') and right-sided
(``interior'') types. By definition, Poincar\'{e} transformations are exterior
ones. We consider all rotations in the seven-dimensional space that (1)
conserve the spacetime components of the particle and antiparticle currents
and (2) do not couple the right-chiral neutrino. These rotations comprise
additional exterior transformations that commute with the Poincar\'{e} group
and form the group $SU(2)_{L}$, interior ones that constitute $SU(3)_{C}$, and
a unique group of coupled double-sided rotations with $U(1)_{Y}$ symmetry. The
spinor mediates a physical coupling of Poincar\'{e} and isotopic symmetries
within the restrictions of the Coleman--Mandula theorem. The four extra
spacelike dimensions in the model form a basis for the Higgs isodoublet field,
whose symmetry requires the chirality of $SU\left(  2\right)  .$ The charge
assignments of both the fundamental fermions and the Higgs boson are produced exactly.

\end{abstract}

%%\eads{\mailto{traylin@uwindsor.ca}, \mailto{baylis@uwindsor.ca}%
%%\\
%%\maketitle

\section{Introduction}

The present work introduces a geometric approach to the minimal standard model
in terms of Clifford's geometric algebra $C\!\ell_{7}$ of seven-dimensional
space (see for example \cite{Hestenes66,Porteous95,Lounesto,Baylis99} for an
introduction to Clifford algebras and their applications in physics). It
demonstrates how the gauge symmetries $U(1)_{Y}\otimes SU(2)_{L}\otimes
SU(3)_{C}$ arise as the rotational symmetries of a reducible representation of
the Poincar\'{e} group in a linear space with only four extra spacelike
dimensions. The fact that this is fewer than the minimum of seven extra
dimensions required in the Kaluza-Klein type of approach \cite{Witten81} stems
both from the availability of double-sided transformations on algebraic spinor
elements and from the existence of higher-dimensional multivector subspaces in
$C\!\ell_{7}$. Our approach of studying rotational symmetries in a
higher-dimensional space may be viewed as an extension of the well-known
association of spin with spatial rotations and the treatment of charge
symmetry as a rotational symmetry in isospin space. It may lead to a better
understanding of the geometry underlying the standard model.

There have been numerous attempts in the past to combine the existing
symmetries into an encompassing structure. Many of the earlier ones have
fallen victim to theorems such as the one by Coleman and Mandula
\cite{Coleman-Mandula} that disallow most except ``trivial'' (i.e.
direct-product) couplings of internal and spacetime symmetries of the $S$
matrix. One of the motivations of supersymmetric models has been to evade the
restrictions of such theorems \cite{Haag75,Kaku,Chaichian}.

More recently Clifford algebras have been used to model the leptons and quarks
and their interactions \cite{CF89,CF92,CF93,Chisholm,CF99}. The Clifford
algebras $C\!\ell_{3}$ of physical space and $C\!\ell_{1,3}$ of Minkowski
spacetime are just the algebra of Pauli spin matrices and the real algebra of
Dirac gamma matrices, respectively, and are essential ingredients of the
relativistic quantum theory of fermions. The aim is to find a larger algebra
containing $C\!\ell_{3}$ and $C\!\ell_{1,3}$ as subalgebras that models
several fermions simultaneously. Chisholm and Farwell \cite{Chisholm}
investigated the mathematical constraints on models in which the particle
spinors belong to minimal left ideals of the algebra, and they developed a
spin-gauge theory \cite{CF89} in which couplings to a ``frame field'' are
responsible for the boson masses. They studied $C\!\ell_{1,6}$ \cite{CF92},
represented by $8\times8$ matrices, to model the electroweak interactions of
leptons in a seven-dimensional space, $C\!\ell_{2,6}$ \cite{CF89}, represented
by $16\times16$ matrices, to combine the electroweak and gravitational
interactions in an eight-dimensional space, and both $C\!\ell_{4,7}$
\cite{CF92} and $C\!\ell_{3,8}$ \cite{CF93}, represented by $32\times32$
matrices, to model one generation of eight fermions (without separate
antiparticles) in an eleven-dimensional space. In all these models, gauge
transformations acted on the spinors only from the left. In 1999 they chose a
different behaviour \cite{CF99} for the gauge transformations in which the
spinors, still taken to be minimal left ideals, undergo similarity
transformations. They interpreted a new interaction term resulting from this
formulation in $C\!\ell_{1,6}$ as representing the $U(1)$ contribution in
electroweak theory.

Our algebraic approach builds on a previous formulation \cite{Baylis1} in
geometric algebra of the Dirac theory. While it shares many of the powerful
tools and algebraic structures of Clifford algebras with the work of Chisholm
and Farwell, it is distinct in several respects. Our spinor, representing all
the fermions and their antiparticles for a single generation of the standard
model, is not an element of a minimal left ideal. However, isotopic pairs of
particles can be isolated in the spinor by applying primitive idempotents on
the right, and such projected spinors do belong to distinct minimal left
ideals. The transformation behaviour is determined by the geometric role of
the spinor \cite{Hestenes66,Baylis1,Baylis1a} as an amplitude of the Lorentz
transformation relating a reference frame for the particles to the lab frame:
the spinor is subject to \emph{independent} transformations on the right and
left. It is through this structure, together with the Minkowskian metric of
paravector space \cite{Baylis99}, that we are able to model all the fermions
of a generation in just seven spatial (eight spacetime) dimensions with an
algebra represented by $8\times8$ matrices. More important than the
compactness of our model, however, are its results. Our work emphasizes the
geometrical significance of the algebra. The spinors physically couple
``interior'' and ``exterior'' symmetries (to be defined below) in such a way
as to maintain a direct-product group structure in their two-sided
transformations. The $SU(2)$ and $SU(3)$ symmetries arise as the exact
exterior and interior rotation groups, respectively, in the seven-dimensional
space that (1) conserve the physical spacetime components of particle and
antiparticle currents and (2) leave the right-chiral neutrino sterile. The
$U(1)$ symmetry is given by coupled rotations that act simultaneously on both
sides of the spinor, commute with the interior and exterior rotations, and
satisfy the constraints (1) and (2). It is important to emphasize that in our
geometric model the gauge symmetries are not imposed but arise naturally from
the algebra itself as unique symmetry groups of the current. The chiral nature
of the $SU\left(  2\right)  $ group is discussed in terms of the symmetry of
the Higgs field. The model also predicts the correct weak-hypercharge assignments.

Section 2 summarizes the conventions adopted and provides an $8\times8$-matrix
representation of $C\!\ell_{7}$ in order to relate our algebraic formulation
to conventional expressions. Section 3 develops the notion of spinors in
$C\!\ell_{7}$. It shows how spinors representing all the fermions of a single
generation can be combined into a single algebraic spinor, how the currents
are calculated from such spinors, and how the contributions from individual
fermions can be projected out. In section 4, we study the rotational
symmetries of these spinors and show that they give exactly the gauge
symmetries of the standard model with the correct weak hypercharge
assignments. We also investigate other possible symmetry transformations and
show that within our model the continuous interior and exterior symmetry
groups (other than the Poincar\'{e} group) comprise only sets of coupled
rotations. Section 5 shows how the four extra spatial dimensions and their
transformation properties are precisely what is needed for the four components
of a minimal Higgs field.

\section{Algebraic Foundations}

Clifford algebras are associative algebras of vectors. In the real Clifford
algebra $C\!\ell_{7}$, the unit vectors $e_{1},e_{2},...,e_{7}$ are chosen to
represent orthogonal spacelike directions in the tangent space of a
seven-dimensional manifold, with $e_{1},e_{2},e_{3}$ allotted to the three
observed (physical) directions. The product of any number of vectors is
completely determined by the anticommutator
\begin{equation}
e_{j}e_{k}+e_{k}e_{j}=2\delta_{jk},\quad j,k=1,\ldots,7. \label{algebra}%
\end{equation}
All elements of the algebra can be reduced to real linear combinations of
$2^{7}=128$ basis forms, each one representing a geometric object. For
example, the bivector $e_{1}e_{4}$ represents the plane spanned by the
directions $e_{1}$ and $e_{4}$, and the trivector $e_{1}e_{2}e_{3}$ represents
the physical volume element. There are a total of 21 independent bivectors and
in general ${{{{{{{{{{{{{{{{{{{{{\binom{7 }{k}}}}}}}}}}}}}}}}}}}}}} $
independent $k $-vectors (forms built from products of $k$ distinct basis
vectors) in $C\!\ell_{7}\,. $

Two basic conjugations, both of which are antiautomorphic involutions, are
used. The reversion of $K\in C\!\ell_{7}$, denoted $K^{\dagger}$, reverses the
order of appearance of all vector elements within $K$. For example,
$(e_{1}e_{2}e_{3})^{\dagger}=e_{3}e_{2}e_{1}=-e_{1}e_{2}e_{3}$ and
$(AB)^{\dagger}=B^{\dagger}A^{\dagger}.$ Clifford conjugation, denoted by
$\bar{K},$ both reverses the order and negates all vector elements of $K$. In
the algebras $C\!\ell_{n},$ the basis vectors $e_{j}$ can all be taken to be
hermitian, and then reversion is equivalent to hermitian conjugation. The
algebra $C\!\ell_{7}$ is appealing in that the volume element of the algebra,
like that of $C\!\ell_{3}\,,$ commutes with all elements and squares to $-1$.
It can therefore be associated identically with the unit imaginary
\begin{equation}
\mathrm{i}\equiv e_{1}e_{2}e_{3}e_{4}e_{5}e_{6}e_{7} \label{volume}%
\end{equation}
and used to reduce products of real vectors to elements of a complex space
with 64 basis forms. For example, $e_{4}e_{5}e_{6}e_{7}=-\mathrm{i}e_{1}%
e_{2}e_{3}$. This fortuitous circumstance occurs for every $C\!\ell_{3+4n}$
with non-negative integer $n$, and $C\!\ell_{7}$ is the smallest of the series
that contains the Dirac algebra as a subalgebra. The choice of adding exactly
four extra dimensions to physical space is further justified below in that
they arise naturally from a metric-free approach to physical space and form a
natural basis for the four components of the minimal Higgs field.

The formalism used here builds on the physical applications of $C\!\ell_{3}$
(the Pauli algebra), in particular the use of \emph{paravectors }%
\cite{Baylis99,Baylis2a} to model spacetime vectors. Paravectors are sums of
scalars and vectors such as $V=V^{0}+V^{1}e_{1}+V^{2}e_{2}+V^{3}e_{3}\equiv
V^{\mu}e_{\mu}$, where for notational convenience we denote the unit scalar by
$e_{0},$ and the scalar $V^{0}$ is the time component in the observer frame,
that is, the frame with proper velocity $e_{0}=\bar{e}_{0}=1$. The linear
space of paravectors has a Minkowski spacetime metric $\eta_{\mu\nu}$ with
signature $(1,3).$ The metric arises from the square norm of paravectors
\begin{equation}
V\bar{V}=\left\langle V\bar{V}\right\rangle _{S}=V^{\mu}V^{\nu}\left\langle
e_{\mu}\bar{e}_{\nu}\right\rangle _{S}=V^{\mu}V_{\mu}%
\end{equation}
as $\eta_{\mu\nu}=\left\langle e_{\mu}\bar{e}_{\nu}\right\rangle _{S}$. Here,
$\left\langle \cdots\right\rangle _{S}$ means the scalar part of the enclosed
expression, and we adopt the summation convention for repeated indices, with
lower-case Greek indices taking integer values $0\ldots3.$ The algebra
generated by products of paravectors is just $C\!\ell_{3},$ which is
isomorphic to quaternions over the complex numbers. It admits a covariant
formulation of relativity and has also been shown to provide a natural
formulation of the single-particle Dirac theory \cite{Baylis1}. The
Lorentz-invariant spacetime volume element in $C\!\ell_{3}$ can be taken to be
$e_{0}e_{1}e_{2}e_{3}=\pm\mathrm{i\,.}$ The sign indicates the handedness of
the spatial basis vectors $\left\{  e_{1},e_{2},e_{3}\right\}  .$ As we
discuss in more detail in the following section, when extra dimensions are
present, it is possible to rotate a right-handed spacetime basis into a
left-handed one.

Proper and orthochronous Lorentz transformations of spacetime vectors are
effected by bilinear transformations of the form \cite{Baylis3}%
\begin{equation}
V\rightarrow LVL^{\dag} \label{LTV}%
\end{equation}
where $L$ is any unimodular element: $L\bar{L}=1.$ Every such $L$ can be
expressed as the product $L=\exp\left(  w/2\right)  \exp\left(  \theta
/2\right)  $ of a spatial rotation $L_{R}=\exp\left(  \theta/2\right)  $ in
the plane of the bivector $\theta=\frac{1}{2}\theta^{jk}e_{j}\bar{e}_{k}$ and
a pure boost $L_{B}=\exp\left(  w/2\right)  $ in the direction of the rapidity
$w=w^{j}e_{j}$ (or, equivalently, as a hyperbolic rotation in the spacetime
plane of $w^{j}e_{j}\bar{e}_{0}$). The scalar coefficients satisfy
$\theta^{jk}=-\theta^{kj}$ and $w^{j}=0=\theta^{kj}$ for $j>3.$ An advantage
of the formalism is that the generators of the transformations have direct
physical significance. For example, the generator $e_{1}\bar{e}_{2}$ induces a
rotation in the $e_{1}e_{2}$ plane. Note that a scalar is not necessarily the
time component of some spacetime vector. The mass $m$ of a particle, for
example, may be the time component of the momentum $p$ (in units with $c=1$)
in the rest frame, or it may be the invariant norm of $p.$ The two
possibilities are distinguished by how they transform. In particular, the
square norm of $p$ transforms as%
\begin{equation}
m^{2}=p\bar{p}\rightarrow\left(  LpL^{\dag}\right)  \left(  \bar{L}^{\dag}%
\bar{p}\bar{L}\right)  =p\bar{p}%
\end{equation}
whereas the rest-frame momentum becomes%
\begin{equation}
me_{0}\rightarrow Lme_{0}L^{\dag}=mLL^{\dag}.
\end{equation}

The extension from $C\!\ell_{3}$ to $C\!\ell_{7}$ requires four additional
basis vectors, $e_{4},e_{5},e_{6},e_{7},$ that are orthogonal to physical
space, namely the span of $\left\{  e_{1},e_{2},e_{3}\right\}  ,$ which
generates the $C\!\ell_{3}$ considered here. If $z$ is any linear combination
of $e_{4},e_{5},e_{6},e_{7},$ its product with any $K\in C\!\ell_{3}$
satisfies%
\begin{equation}
zK=\bar{K}^{\dag}z\,. \label{zK}%
\end{equation}
It follows that $z$ is invariant under any Lorentz transformation (\ref{LTV})
with $L\in C\!\ell_{3}$: $z\rightarrow LzL^{\dag}=L\bar{L}z=z\,.$ More general
rotations in $C\!\ell_{7}$ have the form of equation\thinspace(\ref{LTV}) but
are generated by bivectors that are not restricted to the three spatial planes
of $C\!\ell_{3}\,.$

It is natural to question the significance of the extra dimensions. Of course
they may be compact as in the Kaluza-Klein approach, but then one can still
perform rotations in the tangent space at any point, for example in the
$e_{1}e_{4}$ plane or in other planes involving the extra dimensions.
Alternatively, the extra dimensions may be finite or infinite in extent but
simply not observable as spatial degrees of freedom. One way to arrive at
$C\!\ell_{7}$ from $C\!\ell_{3}$ is to seek a metric-free foundation for
$C\!\ell_{3}.$ The anticommutation relation (\ref{algebra}) implies a
Euclidean spatial metric, but we may instead start with a three-dimensional
metric-free Witt basis \cite{Crumey90,Doran93} of null vectors $\left\{
\alpha_{1},\alpha_{2},\alpha_{3}\right\}  $ satisfying%
\begin{equation}
\alpha_{j}\alpha_{k}+\alpha_{k}\alpha_{j}=0,\quad j,k=1,2,3. \label{anticomm}%
\end{equation}
A dual space can then be defined as the span of $\left\{  \alpha_{1}^{\ast
},\alpha_{2}^{\ast},\alpha_{3}^{\ast}\right\}  ,$ where%
\begin{equation}
\alpha_{j}^{\ast}\alpha_{k}+\alpha_{k}\alpha_{j}^{\ast}=\delta_{jk}\,.
\label{anticom}%
\end{equation}
The anticommutation relation (\ref{algebra}) for $C\!\ell_{3}$ follows
directly from the identification $e_{k}=\alpha_{k}+\alpha_{k}^{\ast}.$
However, there are now three extra linearly independent vectors that we can
label $e_{-k}=\alpha_{k}-\alpha_{k}^{\ast}.$ It is easily verified that the
six basis vectors $e_{\pm k}$ anticommute and square to $\pm1.$ The span of
$\left\{  e_{\pm k}\right\}  _{1\leq k\leq3}$ is a six-dimensional space with
the metric signature $(3,3)$. It generates the Clifford algebra $C\hspace
{-0.8mm}\ell_{3,3}$, and its volume element $e_{4}\equiv e_{-3}e_{-2}%
e_{-1}e_{1}e_{2}e_{3}$ squares to $+1$ and anticommutes with the six $e_{\pm
k}.$ As in the familiar Dirac algebra, the volume element in $C\hspace
{-0.8mm}\ell_{3,3}$ acts as an additional spatial dimension. It can be added
to the basis to form a seven-dimensional space with the corresponding
universal Clifford algebra$\,C\hspace{-0.8mm}\ell_{4,3}.$ The algebra
$C\!\ell_{4,3}$ can be mapped to $C\!\ell_{7}$ if we assume the existence of a
scalar unit imaginary element $\mathrm{i}.$ We replace the three $e_{-k}$ by
elements $e_{4+k}\equiv\mathrm{i}e_{-k}$ that square to $+1.$ The elements
$e_{j},$ with $j=1,2,\ldots,7,$ then satisfy equations\thinspace
(\ref{algebra}) and (\ref{volume}) and span a seven-dimensional Euclidean
space such as used here. The Witt basis elements can now be written%
\begin{equation}
\alpha_{k}=\frac{1}{2}\left(  e_{k}-\mathrm{i}e_{4+k}\right)  ,\quad k=1,2,3
\label{alpha}%
\end{equation}
and if we take the $e_{j}$ to be hermitian, the dual elements are their
hermitian conjugates: $\alpha_{k}^{\ast}=\alpha_{k}^{\dag}.$ The
anticommutation relations (\ref{anticomm},\ref{anticom}) are just those of
fermion annihilation and creation operators, whose products, together with
other constructions analogous to equation\thinspace(\ref{alpha}), can generate
the isotopic groups used below. Here we derive the group generators directly
in terms of bivectors of $C\!\ell_{7}$ by demanding that they avoid
interactions with the right-chiral neutrino and leave the spacetime components
of the particle and antiparticle currents invariant.

To illustrate relations in $C\!\ell_{7}$, it is useful to have an explicit
matrix representation. Such a representation can be built from a $4\times
4$-matrix representation of the familiar Dirac algebra $C\!\ell_{1,3}$, in
which the basis vectors satisfy $\gamma_{\mu}\gamma_{\nu}+\gamma_{\nu}%
\gamma_{\mu}=2\eta_{\mu\nu}.$ However, we note that the unit imaginary is not
part of $C\!\ell_{1,3},$ that the volume element $\mathrm{i}\gamma_{5}%
=\gamma_{0}\gamma_{1}\gamma_{2}\gamma_{3}$ plays the role of an added spatial
dimension, that the $\gamma_{\mu}$ cannot all be hermitian, and that
$\gamma_{0}$ has additional significance in the definition of the conjugate
spinor. A faithful $8\times8$ matrix representation of $C\!\ell_{7}$ can be
expressed in the block-matrix form%
\begin{eqnarray}
\fl  e_{0}  &  =1\leftrightarrow\left(
\begin{array}
[c]{ll}%
1 & 0\\
0 & 1
\end{array}
\right)  ,\quad e_{k}\leftrightarrow\left(
\begin{array}
[c]{cc}%
-\gamma_{0}\gamma_{k} & 0\\
0 & -\gamma_{0}\gamma_{k}%
\end{array}
\right)  ,\quad e_{4}\leftrightarrow\left(
\begin{array}
[c]{cc}%
\mathrm{i}\gamma_{0}\gamma_{5} & 0\\
0 & \mathrm{i}\gamma_{0}\gamma_{5}%
\end{array}
\right) \nonumber\\
\fl  e_{5}  &  \leftrightarrow\left(
\begin{array}
[c]{cc}%
\gamma_{0} & 0\\
0 & -\gamma_{0}%
\end{array}
\right)  ,\quad e_{6}\leftrightarrow\left(
\begin{array}
[c]{cc}%
0 & \gamma_{0}\\
\gamma_{0} & 0
\end{array}
\right)  ,\quad e_{7}\leftrightarrow\left(
\begin{array}
[c]{cc}%
0 & -\mathrm{i}\gamma_{0}\\
\mathrm{i}\gamma_{0} & 0
\end{array}
\right)  \label{matrices}%
\end{eqnarray}
with $k=1,2,3.$ Each basis vector $e_{j}$ is thus represented by a hermitian
matrix. It can be seen that this representation absorbs $\gamma_{0}$ into the
definition of a spatial direction, thus relegating time to the scalar part of
the algebra, and it introduces four extra spacelike dimensions in accordance
with the defining anticommutator (\ref{algebra}) so that $[\mathrm{i}%
]_{8\times8}$ arises naturally through the full volume element. Operations
involving these higher dimensions may now be stated and executed cleanly in
terms of the basis vectors $e_{j}$ without having to appeal to products of
gamma matrices. The representation (\ref{matrices}) is only one of many that
absorb the Dirac algebra into the more mathematically uniform $C\!\ell_{7}$.
In fact, the model can be presented algebraically without reference to
specific matrices, but the representation (\ref{matrices}) is useful for
understanding the spinorial element and for making comparisons to conventional expressions.

\section{Algebraic Spinors and Currents}

Algebraic spinors may be defined as entities that transform under the
restricted Lorentz group not as vectors (\ref{LTV}), but according to the
rule
\begin{equation}
\Psi\rightarrow L\Psi. \label{LTS}%
\end{equation}
They obey a similar transformation law under translations. Spinors are thus
elements of the carrier space of a representation (generally a reducible
representation) of the Poincar\'{e} group. In the $C\!\ell_{3}$ version of the
Dirac theory \cite{Baylis1}, the spinor field $\Psi$, represented by a
$2\times2$ matrix, is an amplitude of the bilinear Lorentz transformation
(\ref{LTV}) relating the reference and laboratory frames of the particle. The
current, in particular, corresponds to the transformation of the rest-frame
time axis: $J=\Psi\Psi^{\dag}.$

To describe one generation of the standard model, we use the algebraic spinor
$\Psi\in C\!\ell_{7}$. It is represented by an $8\times8$ matrix whose columns
contain the spinors for the leptons ($\nu,e$) as well as for three colors of
quarks ($u,d$) and all their antiparticles. Presumably it specifies not only
the motion and orientation of the particles in spacetime, but also in the
space spanned by the extra four dimensions.

The transformations (\ref{LTS}) are preserved by multiplication from the right
by Lorentz-invariant factors, in particular by hermitian idempotent elements
(projectors) that project the spinor $\Psi$ onto left ideals of $C\!\ell_{7}.$
In particular, there are eight independent primitive idempotents that, in
terms of matrices, can each be used to reduce $\Psi$ to a single nonvanishing
column. (Examples are given below and in the Appendix.) Every column of $\Psi$
transforms as a spinor [equation\thinspace(\ref{LTS})], and operators from the
left do not mix the spinors in different columns.

In constructing an algebraic expression for the particle current $J$, we seek
a form that is bilinear in the spinors, transforms as a vector, and satisfies
$J^{\dagger}=J$. The last requirement ensures that the physical components
$J^{\mu}$ of $J$ are real. The simplest solution to this, and the one that we
adopt here, has the same form as found for the Dirac theory in $C\!\ell_{3}$:%
\begin{equation}
J=\Psi\Psi^{\dagger}\rightarrow L\Psi\Psi^{\dagger}L^{\dagger}.
\label{current}%
\end{equation}
A specific component of $J$ may be extracted by contracting it with its
associated direction through
\begin{equation}
J_{\mu}=\langle\Psi\Psi^{\dagger}\bar{e}_{\mu}\rangle_{S}=\langle\Psi
^{\dagger}\bar{e}_{\mu}\Psi\rangle_{S}. \label{component}%
\end{equation}
(We have used the algebraic property $\left\langle AB\right\rangle
_{S}=\left\langle BA\right\rangle _{S}$, whose matrix representation through
$\langle\cdots\rangle_{S}\leftrightarrow\frac{1}{8}tr(\cdots)$ is the familiar
trace theorem $tr(AB)=tr(BA)$.) From the matrix representation for $e_{\mu}$,
we see that the components (\ref{component}) are sums of the conventional
expressions $\lceil\bar{\psi}\gamma_{\mu}\psi\rfloor$ for each fermion and
antifermion, where the delimiters $\lceil\ \cdots\rfloor$ designate prevailing
non-algebraic notation.

It is useful to distinguish transformations acting on the left from others
that act on the right. Those on the left include Lorentz transformations and
rotations in the space of the extra four dimensions. Since they operate on
orthogonal subspaces, rotations in the space spanned by $\left\{  e_{4}%
,e_{5},e_{6},e_{7}\right\}  $ commute with the Lorentz transformations. They
are applied to the spinor after the particles have been given the motion and
orientation described by $\Psi$ and will be called ``exterior''
transformations to represent their position, as in equation\thinspace
(\ref{current}), in transformations of the current $J.$ Transformations
applied from the right will similarly be called ``interior''. They are applied
to the particles in their reference frame, before they acquire the motion and
orientation implied by the spinor. Note that exterior transformations are not
synonymous with \emph{external} transformations, since the extra four
dimensions may relate to properties that are commonly considered to be
\emph{internal}. Exterior transformations mix the components within a single
pair of fermions, whereas interior transformations mix different pairs together.

Primitive idempotents $P(n)$ needed to isolate columns of $\Psi$ can be
constructed from interior products of three pairs of simple projectors
$P_{\pm}=P_{\pm}^{\dagger}=P_{\pm}^{2}=\bar{P}_{\mp}\,,$ where $P_{+}+P_{-}=1$
and $P_{+}P_{-}=0$. From among several equivalent choices, we use the three
mutually commuting projector pairs%
\begin{equation}
P_{\pm3}\equiv\frac{1}{2}(1\pm e_{3}),\;P_{\pm\alpha}\equiv\frac{1}{2}%
(1\pm\mathrm{i}e_{4}e_{5}),\;P_{\pm\beta}\equiv\frac{1}{2}(1\pm\mathrm{i}%
e_{6}e_{7}). \label{projectors}%
\end{equation}
In the Weyl $\gamma$-matrix representation adopted here (see Appendix), the
products $P_{\pm3}P_{\pm\alpha}P_{\pm\beta}$ are simply the eight diagonal
matrices with a single nonvanishing, unit element. For example, $P_{+3}%
P_{+\alpha}P_{-\beta}=\mathrm{diag}[1,0,0,0,0,0,0,0]\equiv P(1)$, and the
first-column spinor may be written $\Psi P(1)$ (see Appendix). Each of the
eight primitive projectors $P(n)$, applied from the right, projects $\Psi$ (or
other elements) onto one of eight minimal left ideals of $C\!\ell_{7}$ and one
of the eight columns of the matrix representation. The $n$th column $\Psi
P(n)$ is identified with a distinct pair of fermions and forms current
elements in equation\ (\ref{current}) only with itself.\footnote{A similar
primitive-idempotent structure for particle doublets was proposed for the
algebra $C\!\ell_{1,6}$ by Chisholm and Farwell \cite{Chisholm,CF92}. However,
in spite of an isomorphism between $C\!\ell_{7}$ and $C\!\ell_{1,6},$ their
restriction to spinors belonging to minmal left ideals allows them to include
only one isotopic pair of particles whereas our spinor contains eight isotopic
pairs of particles and antiparticles. Furthermore, our use of paravectors
provides additional degrees of freedom. Indeed, it corresponds to a 2-to-1
mapping of the larger $C\!\ell_{1,7}$ onto its even subalgebra $C\!\ell
_{1,7}^{+}\simeq C\!\ell_{7}\,.$}

One pair of simple projectors, applied from the right, can be taken to
separate particles from antiparticles. We let this be $P_{\pm3}\,,$ although
this choice is generalized below. Thus, columns $1,4,5,8,$ selected by
$P_{+3},$ are designated for particles and the remaining columns, selected by
$P_{-3},$ contain the antiparticle spinors. Each column holds the spinors for
a fermion doublet, and the projectors for the two isotopic-spin components are
taken to be $P_{\pm\beta}$ applied as an \emph{exterior} operator (from the
left). In the Weyl representation \cite{Kaku}, each four-component spinor in
$\Psi$ is further split into two-component spinors of right and left
chirality. For example, the upper spinor of column one comprises the
nonvanishing components of $P_{-\beta}\Psi P(1):$%
\begin{equation}
\left(
\begin{array}
[c]{c}%
\Psi_{11}\\
\Psi_{21}\\
\Psi_{31}\\
\Psi_{41}%
\end{array}
\right)  \equiv\sqrt{8}\left(
\begin{array}
[c]{c}%
\psi_{0}\\
\psi_{1}\\
\psi_{2}\\
\psi_{3}%
\end{array}
\right)  =\sqrt{8}\lceil\left(
\begin{array}
[c]{c}%
\psi_{R}\\
\psi_{L}%
\end{array}
\right)  \rfloor\,.
\end{equation}
where $\sqrt{8}$ factors are inserted to agree with conventional
normalization. The lower spinor $P_{+\beta}\Psi P\left(  1\right)  $ with the
four nonzero components $\Psi_{51}$ to $\Psi_{81}$ and the other $P_{+3}$
columns are labeled in a similar manner. The $P_{+3}$ (particle) spinors can
be factored explicitly as in table \ref{table1}. The $P_{-3}$ spinors have a
similar form but have been excluded for brevity. Indeed, one need only work
out the algebraic equivalent of the first column, since the remaining $P_{+3}
$ columns are easily obtained by multiplying the first-column spinor from the
right by the elements $e_{5}e_{1},e_{5}e_{6},e_{6}e_{1},$ which shifts it to
columns $4,5,8$ respectively. These algebraic spinors transform under
$\Psi\rightarrow L\Psi$ in the same manner as in the conventional column representation.

\begin{table}[h]
\caption{The algebraic $P_{+3}$ (particle) spinors, where the two-component
Weyl spinors are \emph{algebraic} elements defined by $\psi_{R}=\psi_{0}%
+\psi_{1}e_{1},$\ and $\psi_{L}=\psi_{3}-\psi_{2}e_{1}$ for each particle with
Dirac spinor components $\psi_{0},\psi_{1},\psi_{2},\psi_{3}$.}%
\label{table1}
%
%
%%\begin{indented} \item[]%
\begin{tabular}
[c]{@{}ll}%
\br  lower spinor & upper spinor\\
\mr  $P_{+\beta}\Psi P\left(  1\right)  =\sqrt{8}(\psi_{R}e_{6}e_{5}+\psi
_{L}e_{1}e_{6})P\left(  1\right)  $ & $P_{-\beta}\Psi P\left(  1\right)
=\sqrt{8}(\psi_{R}+\psi_{L}e_{1}e_{5})P\left(  1\right)  $\\
$P_{+\beta}\Psi P\left(  4\right)  =\sqrt{8}(\psi_{R}e_{6}e_{1}+\psi_{L}%
e_{6}e_{5})P\left(  4\right)  $ & $P_{-\beta}\Psi P\left(  4\right)  =\sqrt
{8}(\psi_{R}e_{5}e_{1}+\psi_{L})P\left(  4\right)  $\\
$P_{+\beta}\Psi P\left(  5\right)  =\sqrt{8}(\psi_{R}+\psi_{L}e_{5}%
e_{1})P\left(  5\right)  $ & $P_{-\beta}\Psi P\left(  5\right)  =\sqrt{8}%
(\psi_{R}e_{5}e_{6}+\psi_{L}e_{1}e_{6})P\left(  5\right)  $\\
$P_{+\beta}\Psi P\left(  8\right)  =\sqrt{8}(\psi_{R}e_{1}e_{5}+\psi
_{L})P\left(  8\right)  $ & $P_{-\beta}\Psi P\left(  8\right)  =\sqrt{8}%
(\psi_{R}e_{6}e_{1}+\psi_{L}e_{5}e_{6})P\left(  8\right)  $\\
\br  &
\end{tabular}
%%\end{indented}
\end{table}

The chiral projectors for all fermions in the Weyl representation are the
mutually annihilating exterior operators%
\begin{equation}
P_{R/L}=P_{\bar{L}/\bar{R}}=\frac{1}{2}(1\pm e_{4}e_{5}e_{6}e_{7}).
\label{chirality}%
\end{equation}
By the ``pacwoman'' \cite{Baylis99} property $P_{R/L}=\pm e_{4}e_{5}e_{6}%
e_{7}P_{R/L}=\mp\mathrm{i}e_{1}e_{2}e_{3}P_{R/L}\,,$ these projectors split
$C\!\ell_{7}$ into parts in which the basis elements $e_{1},e_{2},e_{3}$ of
physical space have right and left-handed orientations, respectively. In
particular, since
\begin{equation}
e_{1}e_{2}e_{3}P_{R/L}=\pm\mathrm{i}P_{R/L}%
\end{equation}
the spatial volume element $e_{1}e_{2}e_{3}$ (which is equal to the spacetime
volume element $e_{0}e_{1}e_{2}e_{3}$ in the lab frame) can be
replaced\footnote{A potential conflict between these cases is restricted by
the association of observed quantities such as $J_{\mu}$ (\ref{component})
with the scalar expressions.} by $+\mathrm{i}$ when multiplying $P_{R}$ and
$-\mathrm{i}$ when multiplying $P_{L}.$ Note that the chirality projectors
$P_{R/L}$ commute with all elements of the subalgebra $C\!\ell_{3}$ as well as
with $P_{\pm3},P_{\pm\alpha},P_{\pm\beta}$ and therefore with all the
primitive idempotents $P\left(  n\right)  .$ Furthermore, any element $x$ of
$C\!\ell_{7}$ with an odd number of vector factors from the higher dimensions
$e_{4},e_{5},e_{6},e_{7}$ reverses the chirality: $xP_{R}=P_{L}x\,.$ Such
elements include bivectors such as $e_{3}e_{4}$ that can generate rotations
from a left-handed coordinate system into a right-handed one and \emph{vice
versa}. The chirality of $\Psi$ can thus be flipped by the transformation%
\begin{equation}
\Psi\rightarrow-e_{1}e_{2}e_{3}e_{4}\Psi
\end{equation}
which has the effect of reversing the vector components of the current
(\ref{current}) in the span of $\{e_{1},e_{2},e_{3},e_{4}\}$ while leaving the
components in the span of $\{e_{0},e_{5},e_{6},e_{7}\}$ invariant.

Charge conjugation is realized by the algebraic operation
\begin{equation}
\Psi\rightarrow\Psi_{C}=\mathrm{i}e_{4}\overline{\Psi}^{\dagger}.
\label{conjugation}%
\end{equation}
The combination of the two antiautomorphic involutions obeys the rule
$\overline{(AB)}^{\dagger}=\bar{A}^{\dagger}\bar{B}^{\dagger},$ and the
conjugate of the upper spinor of the first column (see table \ref{table1}),
for example, is%
\begin{equation}
P_{+\beta}\Psi_{C}P\left(  6\right)  =\mathrm{i}e_{4}\sqrt{8}(\bar{\psi}%
_{R}^{\dagger}+\bar{\psi}_{L}^{\dagger}e_{1}e_{5})P(6)
\end{equation}
where $\bar{\psi}_{R}^{\dagger}=\psi_{0}^{\ast}-\psi_{1}^{\ast}e_{1}$ and
$\bar{\psi}_{L}^{\dagger}=\psi_{3}^{\ast}+\psi_{2}^{\ast}e_{1}$. The
identification (\ref{conjugation}), together with the relation (\ref{zK}) and
transformation rule (\ref{LTS}), ensures that spinors and their charge
conjugates transform in the same way under the Lorentz group:
\begin{equation}
\Psi_{C}\rightarrow\mathrm{i}e_{4}\bar{L}^{\dagger}\bar{\Psi}^{\dagger
}=L\mathrm{i}e_{4}\bar{\Psi}^{\dagger}=L\Psi_{C}\,.
\end{equation}
In the matrix representation, charge conjugation (\ref{conjugation}) is
equivalent to defining the conventional charge conjugates through $\lceil
\psi_{C}=\mathrm{i}\gamma^{2}\psi^{\ast}\rfloor$ and interchanging both
corresponding particle and antiparticle columns and upper and lower spinors.
The resulting full structure of $\Psi$ is shown in table \ref{table2}. For the
sake of brevity, we take the liberty of labeling the spinors with the particle
designations shown, although the gauge structure has not yet been determined.
This is one of many possible arrangements and will be generalized below. Note
that charge conjugation reverses the signs on all of the simple interior and
exterior projectors used here.

\begin{table}[h]
\caption{Column designations of the matrix representation of $\Psi$ in terms
of common two-component chiral spinors.}%
\label{table2}
\begin{indented}
\item[]\begin{tabular}{@{}cccccccc}%
\br
$\ell$ & $-\bar{q}_{\mathrm{grn}}$ & $\bar{q}_{\mathrm{blu}%
}$ & $q_{\mathrm{red}}$ &
$q_{\mathrm{grn}}$ & $\bar{\ell}$ & $-\bar{q}%
_{\mathrm{red}}$ & $q_{\mathrm{blu}}%
$\\
\mr
$\nu_{R}$ & $\bar{d}%
_{L}$ & $-\bar{d}_{L}$ & $u_{R}$ & $u_{R}$ &
$-\bar{e}_{L}$ & $\bar{d}%
_{L}$ & $u_{R}$\\
$\nu_{L}$ & $\bar{d}_{R}$ & $-\bar{d}_{R}$ & $u_{L}%
$ & $u_{L}$ &
$-\bar{e}_{R}$ & $\bar{d}_{R}$ & $u_{L}$\\
$e_{R}%
$ & $-\bar{u}_{L}$ & $\bar{u}_{L}$ & $d_{R}$ & $d_{R}$ & $\bar{\nu}%
_{L}$
& $-\bar{u}_{L}$ & $d_{R}$\\
$e_{L}$ & $-\bar{u}_{R}$ & $\bar{u}%
_{R}$ & $d_{L}$ & $d_{L}$ & $\bar{\nu}_{R}$
& $-\bar{u}_{R}$ & $d_{L}%
$\\%
\br
\end{tabular}
\end{indented}
 \end{table}

Geometrically, charge conjugation transforms the particle current as
\begin{equation}
J=\Psi\Psi^{\dagger}\rightarrow e_{4}\bar{\Psi}^{\dagger}\bar{\Psi}e_{4}%
=e_{4}\bar{J}e_{4}%
\end{equation}
and has the effect of negating the $e_{4}$ component while leaving all other
directions invariant. This is a discrete symmetry of the higher-dimensional
directions that is not accessible by a rotation. The negation of two or four
directions can be achieved by rotations, and to negate three directions one
simply reverses one direction followed by reversing another two or four. The
choice of $e_{4}$ and the phase introduced in equation (\ref{conjugation}) are
merely convenient choices for the representation used.

The total current obtained by simply adding all the left ideal doublets into a
single element $\Psi$ is then
\begin{equation}
J\equiv\Psi\Psi^{\dagger}=\sum_{a=1}^{16}\lceil J_{(a)}^{\mu}\rfloor e_{\mu
}+\mbox{(higher-dim.\ terms).}  \label{current2}%
\end{equation}
The sum here runs over the 16 four-component spinors assigned to the upper and
lower halves of the eight minimal left ideals, each of which is ascribed to a
distinct fermion \cite{Chisholm}. The residual part of the current involves
cross-current terms between the upper and lower fermions of the same ideal as
well as mass-like terms of the form $\lceil\bar{\psi}\psi\rfloor$, all
projected onto higher-dimensional elements.

The main idea of this section has simply been that, instead of writing a
separate term for each of the particle currents, we can consolidate them into
a single expression that accommodates a number of spinorial representations.
The advantage of the algebraic formalism becomes evident when we enumerate all
the possible rotational symmetries of this current.

\section{Gauge Symmetries}

The algebraic current (\ref{current}) holds all the chiral currents of a
single generation of the standard model, with distinct antiparticle currents,
as a generalized current in a linear space of seven spatial dimensions. In
this section, we show that rotational transformations that leave both the
physical spacetime components of the particle and antiparticle currents and
the right-chiral neutrino (and left-chiral antineutrino) invariant lead
exactly to the standard-model gauge symmetries. Our approach is analogous to
the conventional case where one notices that $\lceil\psi\rightarrow
\exp(i\theta)\psi\rfloor$ is a symmetry of the current, but now we consider
all possible rotations in the seven-dimensional Euclidean space. This involves
generators acting from both the left and right of the algebraic spinor, as
these generators usually do not commute with $\Psi$. We show further that
rotations are the only continuous transformations acting from either the right
or the left that are allowed in our model. Thus, by combining the fermion
currents into the single form (\ref{current}), we uncover relationships among
the fermions that in most other models are simply imposed on abstract spaces.

We begin by considering exterior rotations $\Psi\rightarrow G\Psi=\exp(\theta
T)\Psi$ that leave the physical spacetime components of $\Psi\Psi^{\dagger}$
invariant, where $T$ generates rotations in one or more planes of the
seven-dimensional space. As seen above, the generator of rotations in a plane
is the bivector for the plane, and bivectors are antihermitian. From the
infinitesimal form
\begin{equation}
J\rightarrow(1+\theta T)\Psi\Psi^{\dagger}(1+\theta T^{\dagger})
\end{equation}
it is clear from the invariance of $J_{\mu}$ (\ref{component}) for
$\mu=0,1,2,3,$ that $e_{\mu}T=-T^{\dagger}e_{\mu}=Te_{\mu}$. Thus, to leave
the spatial components of $J$ invariant, $T$ must commute with $e_{k}$,
$k=1,2,3$. This reduces the choices for $T$ to linear combinations of the six
bivectors $e_{j}e_{k}:(j,k)\in\{4,5,6,7\},\;j>k\,,$ of the Lie algebra
$so\left(  4\right)  ,$ which generate rotations of the higher-dimensional
vector components of the current among themselves. As seen above, generators
formed from products of $e_{4},e_{5},e_{6},e_{7}$ are invariant under Lorentz
transformations and may therefore be associated with intrinsic
transformations. The projectors $P_{R/L}$ split $so\left(  4\right)  $ into
two independent copies of the algebra $su\left(  2\right)  ,$ corresponding to
the rotation groups $SU\left(  2\right)  _{L/R}\,$ with generators of the form
$e_{j}e_{k}P_{L/R}\,.$

The generators of $SU\left(  2\right)  _{L}$ may be written in the form%
\begin{equation}
T_{1}=\frac{1}{4}(e_{6}e_{4}+e_{5}e_{7}),\ T_{2}=\frac{1}{4}(e_{7}e_{4}%
+e_{6}e_{5}),\ T_{3}=\frac{1}{4}(e_{5}e_{4}+e_{7}e_{6}). \label{generators}%
\end{equation}
that implicitly contains the left-chiral projector (\ref{chirality}), for
example $2T_{1}=e_{6}e_{4}P_{L}\,$, and therefore acts only on left-chiral
particles and right-chiral antiparticles. The three generators
(\ref{generators}) induce simultaneous rotations in a pair of commuting planes
and satisfy $[T_{a},T_{b}]=\varepsilon_{abc}T_{c},$ with the fully
antisymmetric structure constants $\varepsilon_{abc}$ where $\varepsilon
_{123}=1.$ The conventional presence of the unit imaginary in front of $T_{c}
$ has been absorbed into the antihermitian property of the bivectors. The
effect of the transformation $\Psi\rightarrow\exp(\theta_{a}T_{a})\Psi$ is
identical to that of the prevailing $SU(2)$ prescriptions
\begin{equation}
\fl  \left(
\begin{array}
[c]{cccc}%
\nu_{L} & u_{L} & -\bar{e}_{R} & -\bar{d}_{R}\\
e_{L} & d_{L} & \bar{\nu}_{R} & \bar{u}_{R}%
\end{array}
\right)  \rightarrow\exp(-\mathrm{i}\theta_{a}\sigma_{a}/2)\left(
\begin{array}
[c]{cccc}%
\nu_{L} & u_{L} & -\bar{e}_{R} & -\bar{d}_{R}\\
e_{L} & d_{L} & \bar{\nu}_{R} & \bar{u}_{R}%
\end{array}
\right)
\end{equation}
as is readily verified by computing the matrix representations of the
generators. Because operations from the left shuffle entire rows about in the
matrix representation but do not shift columns, the assignment of doublets to
columns is still arbitrary. The three linearly independent generators formed
by replacing the $+$ signs in (\ref{generators}) by $-$ signs, and indeed any
linear combination of them, all have the form $x_{b}P_{R},$ where $x_{b}$ is a
bivector. They would thus couple with $\nu_{R}$ and its conjugate and are
therefore omitted.

Now let us look at the possible \emph{interior }rotations $\Psi\rightarrow\Psi
G^{^{\prime}}=\Psi\exp(\theta T^{^{\prime}}).$ To emphasize the fact that they
act on the \emph{right} side of $\Psi$, the interior transformations and
generators are denoted here with a prime. Any interior unitary transformation
leaves $J=\Psi\Psi^{\dag}$ invariant, but we want a stronger condition: we
demand that the spacetime components of the particle and antiparticle currents
be separately invariant. Mathematically, this is equivalent to splitting the
current in two using the $\Psi P_{\pm3}$ spinors%
\begin{equation}
J=\frac{1}{2}\Psi(1+e_{3})\Psi^{\dagger}+\frac{1}{2}\Psi(1-e_{3})\Psi
^{\dagger}\equiv J_{+3}+J_{-3} \label{current3}%
\end{equation}
and requiring each part to be invariant. Recall that $P_{+3}$ and $P_{-3}$ are
projectors for particles and antiparticles, respectively, and remember that
the interior projectors do not Lorentz transform; they represent a choice in
the intrinsic or reference-frame structure of the particles and are not
altered by a Lorentz transformation operating from the opposite side of the
spinors. Generators acting between $\Psi$ and $\Psi^{\dag}$ are similarly
Lorentz invariant. Thus, we may involve the elements $e_{1},e_{2},e_{3}$ in
the interior symmetries while satisfying the Coleman-Mandula theorem
\cite{Coleman-Mandula}, which prohibits any non-trivial combination of the
Poincar\'{e} and isotopic groups. Under the infinitesimal interior
transformation $\Psi\rightarrow\Psi(1+\theta T^{^{\prime}})$, we have
\begin{equation}
J_{\pm3}\rightarrow\frac{1}{2}\Psi(1+\theta T^{^{\prime}})(1\pm e_{3}%
)(1+\theta T^{^{\prime}\dagger})\Psi^{\dagger} \label{internal}%
\end{equation}
which may be viewed as a transformation of the central $P_{\pm3}$ projector.
We see that the space of available bivector generators that leave $e_{3}$
invariant is now spanned by the larger set of 15 bivectors $e_{j}%
e_{k}:(j,k)\in\{1,2,4,5,6,7\},\;j<k.$ Insulating the right-chiral neutrino
from interior transformations in a similar manner as before now requires that
both lepton columns (1 and 6 in the representation adopted) be avoided. This
reduces the number of independent generators to eight, all of which couple
quarks of different colour charges:%
\begin{eqnarray}
T_{1}^{^{\prime}}  &  =\frac{1}{4}(e_{1}e_{7}+e_{6}e_{2}),\ T_{2}^{^{\prime}%
}=\frac{1}{4}(e_{1}e_{6}+e_{2}e_{7}),\ T_{3}^{^{\prime}}=\frac{1}{4}%
(e_{1}e_{2}+e_{7}e_{6})\nonumber\\
T_{4}^{^{\prime}}  &  =\frac{1}{4}(e_{6}e_{4}+e_{5}e_{7}),\ T_{5}^{^{\prime}%
}=\frac{1}{4}(e_{4}e_{7}+e_{5}e_{6}),\ T_{6}^{^{\prime}}=\frac{1}{4}%
(e_{4}e_{1}+e_{2}e_{5})\nonumber\\
T_{7}^{^{\prime}}  &  =\frac{1}{4}(e_{1}e_{5}+e_{2}e_{4}),\ T_{8}^{^{\prime}%
}=\frac{1}{4\sqrt{3}}(e_{2}e_{1}+2e_{5}e_{4}+e_{7}e_{6})\,. \label{Tprime}%
\end{eqnarray}
The interior generators have been arranged to give the conventional $SU(3)$
structure constants \cite{Kaku}%
\begin{equation}
\lbrack T_{a}^{^{\prime}},T_{b}^{^{\prime}}]=-f_{abc}T_{c}^{^{\prime}}\,.
\label{SU3gens}%
\end{equation}

Computing the matrix representation for each of these generators using
(\ref{matrices}), we find that the transformation $\Psi\rightarrow\Psi
\exp(\theta_{a}T_{a}^{^{\prime}})$ is identical in its effect on the $P_{+3}$
spinor components to
\begin{equation}
\left(  q_{\mathrm{red}},q_{\mathrm{grn}},q_{\mathrm{blu}}\right)
\rightarrow\left(  q_{\mathrm{red}},q_{\mathrm{grn}},q_{\mathrm{blu}}\right)
\exp(-\mathrm{i}\theta_{a}\lambda_{a}^{\ast}/2)
\end{equation}
where $\lambda_{a}$ are the Gell-Mann matrices. This is equivalent to the more
familiar
\begin{equation}
\left(
\begin{array}
[c]{l}%
q_{\mathrm{red}}\\
q_{\mathrm{grn}}\\
q_{\mathrm{blu}}%
\end{array}
\right)  \rightarrow\exp(-\mathrm{i}\theta_{a}\lambda_{a}/2)\left(
\begin{array}
[c]{l}%
q_{\mathrm{red}}\\
q_{\mathrm{grn}}\\
q_{\mathrm{blu}}%
\end{array}
\right)  .
\end{equation}
Under the same algebraic operation, the effect of the remaining submatrices on
the conjugate spinors $(-\overline{q_{\mathrm{grn}}},\overline{q_{\mathrm{blu}%
}},-\overline{q_{\mathrm{red}}})$ is equivalent to
\begin{equation}
\left(  \overline{q_{\mathrm{red}}},\overline{q_{\mathrm{grn}}},\overline
{q_{\mathrm{blu}}}\right)  \rightarrow\left(  \overline{q_{\mathrm{red}}%
},\overline{q_{\mathrm{grn}}},\overline{q_{\mathrm{blu}}}\right)
\exp(\mathrm{i}\theta_{a}\lambda_{a}/2)
\end{equation}
which is the correct transformation. The fact that the doublets can be written
in the same representation by using either the column $(u,d)$ or the column
$(-\bar{d},\bar{u})$ is a special property of $SU(2).$ Such a construction is
not possible in the for the $SU(3)$ triplet, but the geometric symmetries here
provide a separate set of $SU(3)$ submatrices, one in terms of $-\lambda
_{a}^{\ast}$ and the other in terms of $\lambda_{a}$, operating on the two
carrier spaces. It is an advantage of having the conjugate spinors in separate
columns of $\Psi,$ that the same algebraic symmetry applies to both particles
and antiparticles.

Since any operation from the left shuffles rows whereas one from the right
shuffles columns, the order in which two such operations is applied is
immaterial. Therefore, \emph{it is of no consequence that the generators from
the left do not necessarily commute with the generators acting from the
right.} They act on independent structural elements (rows and columns) of
$\Psi$ and thus effect transformations as if they were two commuting
symmetries in an abstract space. This property, together with the
higher-dimensionality of the linear subspace of bivectors, is basically how
these gauge groups arise from only four extra dimensions.

There remains one additional possible symmetry. We need to consider a
synchronized double-sided rotation that conspires to cancel out in the case of
the right-chiral neutrino. As this rotation is to represent a distinct
symmetry, its left- and right-side generators must commute with all $SU(2)$
and $SU(3)$ generators, respectively. Since both the right- and left-sided
parts separately couple the right-chiral neutrino, we resurrect previously
discarded generators. The surviving bivector candidates are $(e_{4}e_{5}%
+e_{7}e_{6})$ acting from the left, and $(e_{1}e_{2}+e_{5}e_{4}+e_{6}e_{7})$
operating from the right. One may verify with the infinitesimal operator%
\begin{equation}
\Psi\rightarrow(1+\theta_{0}T_{0})\Psi(1+\theta_{0}T_{0}^{\prime})
\label{synch}%
\end{equation}
that the solution for which there is no change to the right-chiral neutrino
can be normalized to%
\begin{equation}
T_{0}=\frac{1}{2}(e_{4}e_{5}+e_{7}e_{6}),\ T_{0}^{\prime}=\frac{1}{3}%
(e_{1}e_{2}+e_{5}e_{4}+e_{6}e_{7}). \label{beta}%
\end{equation}
Applying this operation to each spinor in turn proves to be identical to the
$U(1)_{Y}$ transformation $\psi_{(j)}\rightarrow\exp(-\mathrm{i}\theta
_{0}Y_{(j)})\psi_{(j)}$ with the weak hypercharge assignments%
\begin{eqnarray}
Y(\nu_{R},\nu_{L},e_{R},e_{L})  &  =(0,-1,-2,-1)=-Y(\bar{\nu}_{L},\bar{\nu
}_{R},\bar{e}_{L},\bar{e}_{R})\\
Y(u_{R},u_{L},d_{R},d_{L})  &  =\left(  4/3,1/3,-2/3,1/3\right)
=-Y(\bar{u}_{L},\bar{u}_{R},\bar{d}_{L},\bar{d}_{R}).\nonumber
\end{eqnarray}
It produces the conventional weak hypercharge assignments for both leptons and quarks.

The above transformations may now be combined into a single expression
\begin{equation}
\Psi\rightarrow\exp(\theta_{0}T_{0}+\theta_{a}T_{a})\Psi\exp(\theta_{0}%
T_{0}^{\prime}+\theta_{b}^{\prime}T_{b}^{\prime})
\end{equation}
operating on both particles and antiparticles. This exhausts the rotational
gauge symmetries. The double-sided transformations may be locally gauged by
introducing twelve gauge fields $B,W_{a},G_{a}\in C\!\ell_{3}$ that transform
according to%

\begin{eqnarray}
\bar{B}  &  \rightarrow\bar{B}+\frac{2}{g^{\prime}}\bar{\partial}\theta
_{0},\nonumber\\
\bar{W}_{a}  &  \rightarrow\bar{W}_{a}+\frac{1}{g}\bar{\partial}\theta
_{a}+\varepsilon_{abc}\theta_{b}\bar{W}_{c}\,,\quad a\in\left\{  1,2,3\right\}
\label{Gtrans}\\
\bar{G}_{a}  &  \rightarrow\bar{G}_{a}+\frac{1}{g_{s}}\bar{\partial}\theta
_{a}^{^{\prime}}+f_{abc}\theta_{b}^{^{\prime}}\bar{G}_{c}\,,\quad a\in\left\{
1,2,\ldots,8\right\} \nonumber
\end{eqnarray}
into the Lagrangian derivative terms
\begin{eqnarray}
\mathcal{L}_{\partial}  &  =\langle\Psi^{\dagger}\mathrm{i}\bar{\partial}%
\Psi\rangle_{S}\nonumber\\
&  -\frac{g^{\prime}}{2}\langle\Psi^{\dagger}\mathrm{i}\bar{B}(T_{0}\Psi+\Psi
T_{0}^{\prime})\rangle_{S}\nonumber\\
&  -g\langle\Psi^{\dagger}\mathrm{i}\bar{W}_{a}T_{a}\Psi\rangle_{S}\nonumber\\
&  -g_{s}\langle\Psi^{\dagger}\mathrm{i}\bar{G}_{a}\Psi T_{a}^{\prime}%
\rangle_{S} \label{L-der}%
\end{eqnarray}
where the algebraic derivative operator is defined by \cite{Baylis99}
\begin{equation}
\bar{\partial}\equiv\partial_{0}+\partial_{1}e_{1}+\partial_{2}e_{2}%
+\partial_{3}e_{3}. \label{derivative}%
\end{equation}
When used with the interior and exterior generators found above, expression
(\ref{L-der}) yields all the usual particle and antiparticle charge currents.
Note that all bivector generators uniformly obey $T^{\dagger}=\bar{T}=-T,$ and
all exterior $T$ commute with the physical gauge fields. Although the above
terms are similar to the conventional forms, it should be emphasized that all
of the currents are simultaneously handled in the same expression using the
algebraic spinor $\Psi$, whose gauge symmetries arise naturally from the
geometry of the model.

It is of interest to relax the condition that the transformations of $\Psi$ be
rotations and to see whether generators other than bivectors might play a
role. However, the unitarity of the transformations together with the
consistency of charge conjugation (\ref{conjugation}) combine with the
invariance of spacetime components of the particle and antiparticle currents
and the sterility of $\nu_{R}$ to restrict both exterior and interior
generators to bivectors. Explicitly, unitarity requires $T$ and $T^{\prime}$
to be antihermitian ($T=-T^{\dagger}$), restricting them to real linear
combinations of products of 2, 3, 6, or 7 vectors. Consistency requires charge
conjugation to commute with the interior and exterior transformations,
yielding $Te_{4}=e_{4}\bar{T}^{\dagger}$ and $T^{\prime}=\bar{T}%
^{\prime\dagger}$. These relations eliminate all odd elements except
trivectors of $T$ that anticommute with $e_{4}$. The invariance of the
$e_{1},e_{2}$, and $e_{3}$ components of $J$ further eliminates 6-vectors as
well as all trivectors except $e_{1}e_{2}e_{3}$ from admissible contributions
to $T$. The separation of particle and antiparticle currents requires
$T^{\prime}$ to commute with $e_{3},$ which reduces the possible contributions
to $T^{\prime}$ to bivectors plus the one 6-vector $ie_{3}$. The only
remaining candidates that are not bivectors are thus $e_{1}e_{2}e_{3}$ for $T$
and $ie_{3}$ for $T^{\prime}$, and both of these can be eliminated because of
their coupling to $\nu_{R}$. Thus, even after relaxing the condition that the
transformations be rotations, we find that the generators of the interior and
exterior transformations must be bivectors. Furthermore, as seen above, in
order to avoid $\nu_{R}$, the exterior generators must belong to the left
ideal in which $T=TP_{L}$, where $P_{L}$ is the simple chiral projector. From
the form (\ref{chirality}) of $P_{L}$, we see that the generators $T$ are
linear combinations of pairs of commuting bivectors, generating simultaneous
rotations in orthogonal planes. Similarly, the interior generators belong to
the right ideal in which $T^{\prime}=P_{q\bar{q}}T^{\prime}$ where
$P_{q\bar{q}}=1-P(1)-P(6)$ (see the Appendix) is the quark-antiquark
projector, which can also be expressed as a linear combination of simple
projectors. The generators $T^{\prime}$ are thus also seen to be linear
combinations of pairs of commuting bivectors.

A similar relaxation of the rotation requirement for the synchronized
double-sided transformation (\ref{synch}) leaves $T_{0}$ unchanged but adds a
6-vector term to $T_{0}^{\prime}$ (\ref{beta}):
\begin{equation}
T_{0}^{\prime}=\beta(e_{1}e_{2}+e_{5}e_{4}+e_{6}e_{7})+(1-3\beta)ie_{3}\,.
\end{equation}
By restricting the possible transformations to rotations in the
seven-dimensional space, we have effectively chosen $\beta=1/3$.

The $U(1)_{Y}\otimes SU(2)_{L}\otimes SU(3)_{C}$ result here is a general
consequence of the algebra for rotations that conserve the particle and
antiparticle currents and do not couple $\nu_{R}$. They are not specific to
the $\Psi P_{+3}$ spinors. Any arbitrary fitting of the doublets into some
orthogonal linear combination of the columns is accessible by shuffling the
$P_{+3}$ columns through a transformation $\Psi\rightarrow\Psi S$ where
$SS^{\dagger}=1$. The exterior transformations are not effected by this
transformation. The constraint that the transformations are consistent with
charge conjugation demands that $\Psi$ and $\bar{\Psi}^{\dag}$ transform in
the same way, and this implies that $S$ is an even element, comprising only
terms with products of an even number of vectors. The accompanying similarity
transformations $T_{a}^{^{\prime}}\rightarrow S^{\dagger}T_{a}^{^{\prime}}S$
and $P_{+3}\rightarrow S^{\dagger}P_{+3}S$ maintain the results, preserving
the structure constants of the group algebras. It can also be shown that for
any other set in which all the interior generators are written solely as
bivectors, the same weak hypercharge assignments are obtained. In brief, if
$T_{1}^{\prime}$ through $T_{8}^{\prime}$ of the new set are all bivectors, it
can be shown that $S$ must be generated by bivectors. Such a bivector
transformation $S$ on equation\thinspace(\ref{beta}) maintains the same form.
This framework then gives a geometric basis for the gauge group of the
standard model, which arises unambiguously through the various rotational
symmetries of the algebraic current in seven-dimensional space.

\section{Higgs Field}

When looking at the exterior invariances of the current, we previously
disregarded the higher-dimensional vector components and allowed them to
freely rotate among each other. This Lorentz-invariant vector space is then a
carrier space for the set of exterior gauge transformations and affords a
natural inclusion of the minimal Higgs field \cite{Higgs}. With the help of
the matrix representation (\ref{matrices}), one can verify that by formulating
the complex scalar isodoublet $H$ and conjugate Higgs $H_{c}=\bar{H}^{\dagger
}$ as
\begin{eqnarray}
H  &  =(-\phi_{1}e_{6}+\phi_{2}e_{7})P_{-\alpha}+(\phi_{3}e_{5}-\phi_{4}%
e_{4})P_{+\beta}\nonumber\\
&  \sim\left\lceil {{{{{{{{{{{{{{{{{{{{{\binom{\phi_{1}+i\phi_{2} }{\phi
_{3}+i\phi_{4}}}}}}}}}}}}}}}}}}}}}}}\right\rfloor \nonumber\\
H_{c}  &  =(\phi_{1}e_{6}-\phi_{2}e_{7})P_{+\alpha}-(\phi_{3}e_{5}-\phi
_{4}e_{4})P_{-\beta}\nonumber\\
&  \sim\left\lceil {{{{{{{{{{{{{{{{{{{{{\binom{\phi_{3}-i\phi_{4} }%
{\;-\phi_{1}+i\phi_{2}}}}}}}}}}}}}}}}}}}}}}}\right\rfloor \,
\label{Higgsfield}%
\end{eqnarray}
where the $\phi_{j}$ are real scalars, the expression
\begin{equation}
\mathcal{L}_{M}=\frac{1}{\sqrt{2}}\langle\Psi^{\dagger}G_{e}H\Psi P_{\ell
}+\Psi^{\dagger}(G_{d}H+G_{u}H_{c})\Psi P_{q}\rangle_{S} \label{LHiggs}%
\end{equation}
is identical to the conventional Higgs-coupling Lagrangian term with coupling
strengths $G_{e,d,u}.$ The projectors $P_{\ell}=P(1)$ and $P_{q}%
=P(4)+P(5)+P(8)$ are used to separate the lepton and quark currents. The
transformation required for gauge invariance,
\begin{equation}
H\rightarrow\exp(\theta_{0}T_{0}+\theta_{a}T_{a})H\exp(-\theta_{0}T_{0}%
-\theta_{b}T_{b}) \label{Higgstransf}%
\end{equation}
is equivalent to the conventional notation
\begin{equation}
{{{{{{{{{{{{{{{{{{{{{\binom{\phi^{+} }{\phi^{0}}}}}}}}}}}}}}}}}}}}}}%
}\rightarrow\exp(-\mathrm{i}Y\theta_{0}-\mathrm{i}\theta_{a}\sigma
_{a}/2){{{{{{{{{{{{{{{{{{{{{\binom{\phi^{+} }{\phi^{0}}}}}}}}}}}}}}}}}}}}}}}%
\end{equation}
where $\phi^{+}\equiv\phi_{1}+\mathrm{i}\phi_{2}$ and $\phi^{0}=\phi
_{3}+\mathrm{i}\phi_{4}\,.$ The weak hypercharge assignment of $Y=1$ $(Y=-1) $
for the Higgs field (conjugate field) has been recovered naturally from the
double-sided algebraic transformation.

Note that $H$ and $H_{c}$ consist only of odd elements in the span of
$\left\{  e_{4},e_{5},e_{6},e_{7}\right\}  $ and therefore change the
chirality, for example $P_{R}H=HP_{L}\,.$ In fact, they exhaust the couplings
between $R$ and $L$ leptons and between the $R$ and $L$ quarks. Application of
the gauge transformation (\ref{Higgstransf}) using only the exterior
generators of $SU\left(  2\right)  _{L}$ and $U\left(  1\right)  $ naturally
separates the higher-dimensional vector space into the two invariant carrier
spaces of $H$ and $H_{c}$\thinspace and ensures that the gauge fixing occurs
consistently, reducing both $H$ and $H_{c}$ to a component along the same
higher dimension. (We have essentially pre-aligned the Higgs with its
conjugate field by using the same components for each.) On the other hand, the
two generators of rotations that we excluded from possible gauge groups
invariably mix the spaces of $H$ and $H_{c}$. The symmetry of the Higgs
coupling therefore requires the chiral asymmetry seen in the electroweak
interaction. Ignoring the various weighted projectors used in both the Higgs
field (\ref{Higgsfield}) and Lagrangian term (\ref{LHiggs}) to give distinct
masses to the different fermions, the form of equation\thinspace(\ref{LHiggs})
is the same as that of the current\thinspace(\ref{component}), but where the
components of the current being extracted are from the set $\{e_{4}%
,e_{5},e_{6},e_{7}\}$. The Higgs field---one of the least understood aspects
of the standard model---thus arises here simply as a coupling to the
higher-dimensional vector components of the current.

\section{Conclusion}

We began by formulating a generalized current expression in the Clifford
algebra $C\!\ell_{7}$ of seven-dimensional space. The addition of four
spacelike dimensions to those of physical space is the minimum necessary to
incorporate all the fermions of one generation, both particles and
antiparticles, into a single spinorial element. By examining all possible
rotations of the generalized current that leave the right-chiral neutrino (and
left-chiral antineutrino) sterile and conserve the spacetime components of the
particle and antiparticle currents, we found that they are precisely those of
the gauge group of the standard model. The standard-model gauge symmetries are
thus seen to be local rotation groups in the tangent space of a manifold with
\emph{only four} extra spacelike dimensions. In particular, the $SU(2)_{L}$
symmetries arise as \emph{exterior} (left-sided) transformations representing
rotations within the extra dimensions that act only on left-chiral fermions
and their antiparticles, whereas the $SU(3)_{C}$ are \emph{interior}
(right-sided) ones that mix the colour charges of the quarks. The $U(1)_{Y}$
symmetry, complete with all the correct weak hypercharge assignments, arises
as a unique group of double-sided rotations. All of these symmetries commute
with the Poincar\'{e} group and are generated by bivectors of $C\!\ell_{7}$.
The maximal chiral asymmetry of $SU\left(  2\right)  _{L}$ is required by the
symmetry of the Higgs field.

We have further shown that in our model, rotations are in fact the only
continuous transformations allowed that act entirely from the right or from
the left. While we explained our model with the aid of a specific $8\times8$
matrix representation, the symmetries and weak hypercharge assignments depend
only on the algebra and not on any particular selection of primitive
idempotents or ordering of particle spinors in $\Psi$. Finally, the four extra
dimensions required to fit a generation of fermions into a single algebraic
spinor, together with their exterior transformation properties, are precisely
what is needed for the four components of a minimal scalar Higgs field, again
with the correct weak hypercharge assignment.

Many features of the standard model thus flow from the relatively simple
geometry of seven-dimensional Euclidean space, but there are many other
features that call for explanation, such as the origin of the three
generations and the mass spectra. Work is continuing on these and other
aspects of the standard model within the framework of our model.

\subsection*{Acknowledgement}

One of us (G T) would like to thank H Georgi for suggestions on how to make
the Clifford-algebra approach more accessible. This work was supported by the
Natural Sciences and Engineering Research Council of Canada.

\appendix

\section*{Appendix}

\setcounter{section}{1}

Relations for the matrix representation adopted in this paper are summarized
here. There are several versions of the Weyl representation. In this work we
have used
\begin{equation}
\fl  \gamma_{0}=\left(
\begin{array}
[c]{rr}%
0 & -1\\
-1 & 0
\end{array}
\right)  ,\;\gamma_{k}=-\gamma^{k}=\left(
\begin{array}
[c]{cc}%
0 & -\sigma_{k}\\
\sigma_{k} & 0
\end{array}
\right)  ,\newline \gamma_{5}\equiv-i\gamma_{0}\gamma_{1}\gamma_{2}\gamma
_{3},\;
\end{equation}
with%
\begin{equation}
\psi^{Weyl}=\lceil\left(
\begin{array}
[c]{c}%
\psi_{R}\\
\psi_{L}%
\end{array}
\right)  \rfloor\,.
\end{equation}
This is consistent with reference \cite{Kaku}.

The primitive projectors $P\left(  n\right)  $, which are represented by
matrices with elements $P\left(  n\right)  _{jk}=\delta_{jn}\delta_{kn}$, are
given as products of simple commuting projectors (\ref{projectors}) by
\begin{eqnarray}
P\left(  1\right)   &  =P_{+3}P_{+\alpha}P_{-\beta}=\bar{P}\left(  6\right)
\nonumber\\
P\left(  2\right)   &  =P_{-3}P_{+\alpha}P_{-\beta}=\bar{P}\left(  5\right)
\nonumber\\
P\left(  3\right)   &  =P_{-3}P_{-\alpha}P_{-\beta}=\bar{P}\left(  8\right)
\nonumber\\
P\left(  4\right)   &  =P_{+3}P_{-\alpha}P_{-\beta}=\bar{P}\left(  7\right)
\nonumber\\
P\left(  5\right)   &  =P_{+3}P_{-\alpha}P_{+\beta}=\bar{P}\left(  2\right) \\
P\left(  6\right)   &  =P_{-3}P_{-\alpha}P_{+\beta}=\bar{P}\left(  1\right)
\nonumber\\
P\left(  7\right)   &  =P_{-3}P_{+\alpha}P_{+\beta}=\bar{P}\left(  4\right)
\nonumber\\
P\left(  8\right)   &  =P_{+3}P_{+\alpha}P_{+\beta}=\bar{P}\left(  3\right)
\,.\nonumber
\end{eqnarray}
Inverse relations, such as
\begin{equation}
P_{+3}=P\left(  1\right)  +P\left(  4\right)  +P\left(  5\right)  +P\left(
8\right)  \,,
\end{equation}
are easily obtained by summing and applying the complementarity of simple
operators of opposite signs: $P_{+}+P_{-}=1$.

\end{document}